# Optimization Parameter Design of a Circular e+e- Higgs Factory

D. Wang (王逗)[1)], J. Gao (高杰), M. Xiao (肖铭), H. Geng (耿会平), Y. Guo (郭媛媛), S. Xu (许守彦),

N. Wang (王娜), Y. An (安宇文), Q. Qin (秦庆), G. Xu (徐刚), S. Wang (王生)

*Institute of High Energy Physics (IHEP), Beijing 100049, China*

**Abstract**: In this paper we will show a general method of how to make an optimized parameter design of a circular e+e- Higgs Factory by using analytical expression of maximum beam-beam parameter and beamstrahlung beam lifetime started from given design goal and technical limitations. A parameter space has been explored. Based on beam parameters scan and RF parameters scan, a set of optimized parameter designs for 50 km Circular Higgs Factory (CHF) with different RF frequency was proposed.
**Key words**: Circular Higgs Factory (CHF), parameter design, optimization, RF technology
**PACS**: 29.20.db

## 1. Introduction

With the discovery of a Higgs boson on LHC at the energy of about 125 GeV [1, 2], the world high-energy physics community is investigating the feasibility of a Higgs Factory, a complement to the LHC for studying the Higgs. The low Higgs mass makes a circular Higgs Factory possible. Compared with the linear collider, the circular collier as a Higgs Factory has mature technology and rich experience. Also, circular Higgs Factory has potentially a higher luminosity to cost ratio than a linear one at 240 GeV [3]. So, much attention is given to the design of circular Higgs Factory and several proposals have recently been put forward [4-8]. In order to find the optimized machine parameter design started from the required luminosity goal, beam energy, physical constraints at IP and some technical limitations, we study a general analytical method for the parameter choice based on the maximum beam-beam tune shift, beamstrahlung-driven lifetime and beamstrahlung energy spread.

## 2. Beam-beam parameter limit coming from beam emittance blow-up

In e+ e- storage ring colliders, due to strong quantum excitation and synchrotron damping effects, the particles are confined inside a bunch. The position for each particle is random and the state of the particles can be regarded as a gas, where the positions of the particles follow statistic laws. Apparently, the synchrotron radiation is the main source of heating. Besides, when two bunches undergo collision at an interaction point (IP), every particle in each bunch will feel the deflected electromagnetic field of the opposite bunch and the particles will suffer from additional heatings. With the increase of the bunch particle population $N_e$, this kind of heating effect will get stronger and the beam emittance will increase. There is a limit condition beyond which the beam emittance will blow up. This emittance blow-up mechanism introduce a limit for beam-beam tune shift [9]

$$\xi_y \leq \frac{2845}{2\pi}\sqrt{\frac{T_0}{\tau_y \gamma N_{IP}}} \qquad (1)$$

Where $N_{IP}$ is the number of interaction point (When there are $N_{IP}$ interaction points, the independent heating effects have to be added in a statistical way), $\tau_y$ is the transverse damping time and $T_0$ is the revolution time.

## 3. Beam lifetime limit and energy spread limit due to beamstrahlung

When two head-on colliding electron and positron beams penetrate each other, every particle in each beam will feel the electromagnetic field of the other beam and will be defected. This deflection process has some undesirable effects. Firstly, the deflected particle will lose part of its energy due to the synchrotron radiation, called as beamstrahlung, which will increase the energy spread of the colliding beams, and hence increase the uncertainty of the physical experiments. If the beamstrahung is so strong that particles' energy after collision is beyond the

---

ring's energy acceptance, they may leave the beam and strike the vacuum chamber's walls, and hence beam lifetime is decreased. Secondly, the deflected particles will emit photons, hadrons, etc., which will increase the noise background level in the detector. Additionally, after the collision particles will change their flying direction with respect to the axis by a certain angle. If this angle is large enough the particles after the collision will interfere with the detection of small-angle events.

In order to control the extra energy spread by beamstrahlung to a certain degree, we introduce a constraint in this paper as

$$\delta_{BS} \leq \frac{1}{3}\delta_0 \qquad (2)$$

where $\delta_0$ is the nature energy spread and $\delta_{BS}$ is the extra energy spread due to beamstrahlung.

V. I. Telnov [10] pointed out that at energy-frontier e+e− storage ring colliders, beamstrahlung determines the beam lifetime through the emission of single photons in the tail of the beamstrahlung spectra. Unlike the linear collider case, the long tails of the beamstrahlung energy loss spectrum are not a problem because beams are used only once. If we want to achieve a reasonable beamstrahlung-driven beam lifetime of at least 30 minutes, we need to confine the relation of the bunch population and the beam size as follows [7, 11]

$$\frac{N_e}{\sigma_x^* \sigma_z} \leq 0.1\eta \frac{\alpha}{3\gamma r_e^2} \qquad (3)$$

where $\sigma_x^*$ and $\sigma_z$ are the horizontal and longitudinal beam size at IP, $r_e$ is the electron classical radius (2.818×10$^{-15}$m), $\eta$ is the energy acceptance of the ring and $\alpha$ is the fine structure constant (1/137).

4. **Beam parameters calculation**

The luminosity of circular collider is expressed by

$$L[cm^{-2}s^{-1}] = 2.17 \times 10^{34}(1+r)\xi_y \frac{eE_0(GeV)N_bN_e}{T_0\beta_y^*(cm)} F_h \qquad (4)$$

where $r=\sigma_y^*/\sigma_x^*$ is the aspect ratio of the bunch, $\beta_y^*$ is the beta function value at the interaction point, $\xi_y$ is the vertical beam-beam tune shift and $F_h$ is the luminosity reduction factor due to hour glass effect which is expressed as follows

$$F_h = \frac{\beta_y^*}{\sqrt{\pi}\sigma_z} \exp\left(\frac{\beta_y^{*2}}{2\sigma_z^2}\right) K_0\left(\frac{\beta_y^{*2}}{2\sigma_z^2}\right) \qquad (5)$$

where $K_0$ is the zero order modified Bessel function of the second kind.

From eqs. (1) and (4) one finds a limit for the luminosity

$$L_0[cm^{-2}s^{-1}] = 0.7 \times 10^{34}(1+r)\frac{1}{\beta_y^*[cm]}\sqrt{\frac{E_0[GeV]I_b[mA]P_0[MW]}{\gamma N_{IP}}} \qquad (6)$$

$$L_{max} = L_0 F_h \qquad (7)$$

In our method, the goal peak luminosity $L_0$, the energy of the ring $E_0$, the bending radius of the main dipole magnets $\rho$, the synchrotron radiation power $P_0$ (machine technical constraint), the aspect ratio $r$ and the IP number $N_{IP}$ are the known quantity. From these input parameters one gets first

$$U_0 = 88.5 \times 10^3 \frac{E_0^4(GeV)}{\rho} \qquad (8)$$

$$I_b = \frac{P_0}{U_0} \qquad (9)$$

$$\delta_0 = \gamma\sqrt{\frac{C_q}{J_\varepsilon \rho}} \qquad (10)$$

where $U_0$ is the energy loss per revolution due to synchrotron radiation, $I_b$ is the average beam current, $C_q$=3.832×10$^{-13}$ m is a constant and $J_\varepsilon$ is the longitudinal damping partition number (In general case, $J_\varepsilon$=2.).

Then the vertical beta function at IP can be got according to eq. (6)

$$\beta_y^* = \frac{0.7 \times 10^{34}(1+r)}{L_0}\sqrt{\frac{E_0 I_b P_0}{\gamma N_{IP}}} \qquad (11)$$

And the maximum beam-beam tune shift is [9]

$$\xi_y = \frac{2845}{2\pi}\sqrt{\frac{T_0}{\tau_y \gamma N_{IP}}} = \frac{2845}{2\pi}\sqrt{\frac{U_0}{2\gamma E_0 N_{IP}}} \qquad (12)$$

Recalling the original definition of the beam-beam tune shift, for the flat beam, it can be expressed by

$$\xi_y = \frac{N_e r_e \beta_y^*}{2\pi\gamma\sigma_x^*\sigma_y^*} \qquad (13)$$

where $\sigma_x^*$ and $\sigma_y^*$ are the bunch transverse dimensions after the plasma pinch effect (two colliding bunches are fully overlapped).

From eq. (13), one finds

$$\frac{N_e}{\sigma_x^*\sigma_y^*} = \frac{2\pi\gamma}{r_e\beta_y^*}\xi_y \qquad (14)$$

Combining eq. (3) with eq. (14), one has

$$\frac{N_e^2}{\sigma_x^2 \sigma_y \sigma_z} = \frac{0.2\pi\eta\alpha\xi_y}{3r_e^3 \beta_y^*} \tag{15}$$

From the constraint of beamstrahlung energy spread in eq. (2), one finds

$$\frac{N_e^2}{\sigma_x \sigma_y \sigma_z} = \frac{\delta_0}{2.6 r_e^3 \gamma r} \tag{16}$$

So, according to eq. (15) and (16), we get

$$\sigma_x = \frac{5.77 \delta_0 \beta_y^*}{\pi\eta\alpha\xi_y \gamma r} \tag{17}$$

With certain given coupling factor $\kappa_\varepsilon$ (0.005 for example) and the aspect ratio $r$, one can get the vertical beam size and horizontal emittance:

$$\sigma_y = r\sigma_x \tag{18}$$

$$\varepsilon_y = \frac{\sigma_y^2}{\beta_y^*} \tag{19}$$

$$\varepsilon_x = \frac{\varepsilon_y}{\kappa_\varepsilon} \tag{20}$$

$$\beta_x^* = \frac{\sigma_x^2}{\varepsilon_x} \tag{21}$$

From eq. (13) one gets

$$N_e = \frac{2\pi\gamma\xi_y}{r_e \beta_y^*} \sigma_x \sigma_y \tag{22}$$

And also from eq. (3) one gets

$$\sigma_z = \frac{3\gamma r_e^2 N_e}{0.1\eta\alpha\sigma_x} \tag{23}$$

Finally, in order to calculate the total bunch number, we have to refer the expression for the average current

$$I_b = \frac{eN_e N_b}{T_0} \quad (T_0 = \frac{C_0}{c}) \tag{24}$$

where $T_0$ is the revolution time which is decided by the circumference of the ring $C_0$.

Then, having the bunch population eq. (22), it's easy to get the bunch number

$$N_b = \frac{I_b T_0}{eN_e} \tag{25}$$

As a summary, we obtain a set of machine parameters with luminosity goal $L_0$, beam energy $E_0$, ring circumference $C_0$, IP numbers $N_{IP}$, bending radius $\rho$, synchrotron radiation power $P_0$, aspect ratio $r$, coupling factor $\kappa_\varepsilon$ and energy acceptance $\eta$ as input.

$$U_0 = 88.5 \times 10^3 \frac{E_0^4(GeV)}{\rho} \tag{26}$$

$$I_b = \frac{P_0}{U_0} \tag{27}$$

$$\delta_0 = \gamma \sqrt{\frac{C_q}{J_\varepsilon \rho}} \tag{28}$$

$$\xi_{y,\max} = \frac{2845}{2\pi} \sqrt{\frac{U_0}{2\gamma E_0 N_{IP}}} \tag{29}$$

$$\beta_y^* = \frac{0.7 \times 10^{34}(1+r)}{L_0} \sqrt{\frac{E_0 I_b P_0}{\gamma N_{IP}}} \tag{30}$$

$$\sigma_x = \frac{5.77 \delta_0 \beta_y^*}{\pi\eta\alpha\xi_y \gamma r} \tag{31}$$

$$\sigma_y = r\sigma_x \tag{32}$$

$$\varepsilon_y = \frac{\sigma_y^2}{\beta_y^*} \tag{33}$$

$$\varepsilon_x = \frac{\varepsilon_y}{\kappa_\varepsilon} \tag{34}$$

$$\beta_x^* = \frac{\sigma_x^2}{\varepsilon_x} \tag{35}$$

$$N_e = \frac{2\pi\gamma\xi_y}{r_e \beta_y^*} \sigma_x \sigma_y \tag{36}$$

$$\sigma_z = \frac{3\gamma r_e^2 N_e}{0.1\eta\alpha\sigma_x} \tag{37}$$

$$F_h = \frac{\beta_y^*}{\sqrt{\pi}\sigma_z} \exp\left(\frac{\beta_y^{*2}}{2\sigma_z^2}\right) K_0\left(\frac{\beta_y^{*2}}{2\sigma_z^2}\right) \tag{38}$$

$$N_b = \frac{I_b T_0}{eN_e} \tag{39}$$

$$L = L_0 F_h \tag{40}$$

## 5. Optimized design for a 50 km Higgs Factory

### 5.1. Parameter scan

Using the method above, we scan the goal luminosity ($L_0$) with different bending radius $\rho$, IP number $N_{IP}$ and energy acceptance $\eta$ (All the input parameters including the peak luminosity and the technical limitation are listed in table 1). We get some meaningful results which are shown From Fig. 1 to Fig.8. Fig. 1 shows that larger luminosity needs smaller vertical IP beta function, and larger bending radius and less interaction point can lose the IP beta, while the energy acceptance will not affect IP vertical beta function. Fig. 2 shows smaller bending radius and less interaction point give larger vertical beam-beam tune shift, while the parameter $\xi_y$ has no relation with peak luminosity and energy acceptance. Fig. 3 shows that larger luminosity needs smaller bunch population and larger energy acceptance will decrease the bunch population, while the interaction number and bending radius will not affect the bunch population. Fig. 4 tells us that we need more bunch number to get higher luminosity, and also smaller bending radius and smaller energy acceptance can reduce the total bunch number. Meanwhile the bunch number has no relation with the IP number. For the single ring collider, bunch number should not be too large due to the parasitic beam-beam effect. Fig. 5 shows that higher luminosity indicate smaller horizontal emittance (a few nanometer) which suggest a difficulty to design the low emittance lattice with much high energy of 120 GeV (same conclusion as [7]). Also we see that larger bending radius, more IP and smaller energy acceptance will relax the limit for emittance. Fig. 6 tells us that the bunch length has no relation with the peak luminosity and IP number. While smaller bending radius and smaller energy acceptance can help to increase the bunch length. Finally, Fig. 7 and Fig. 8 shows less interaction point, larger bending radius and larger energy acceptance produce larger hour glass factor and hence larger luminosity. So if we want to increase the luminosity we have to increase the bending radius and energy acceptance while reduce the IP number.

TABLE I. Input parameters for machine design

| Energy $E_0$ | Circumference $C_0$ | Goal luminosity $L_0$ | IP number $N_{IP}$ | SR power /beam $P_0$ | Bending radius $\rho$ | aspect ratio $r$ | Coupling $\kappa_\varepsilon$ | Energy acceptance $\eta$ |
|---|---|---|---|---|---|---|---|---|
| 120GeV | 50 km | 1~6×10$^{34}$ cm$^{-2}$s$^{-1}$ | 1~2 | 50 MW | 5~6.2 km | 200 | 0.005 | 5%~12% |

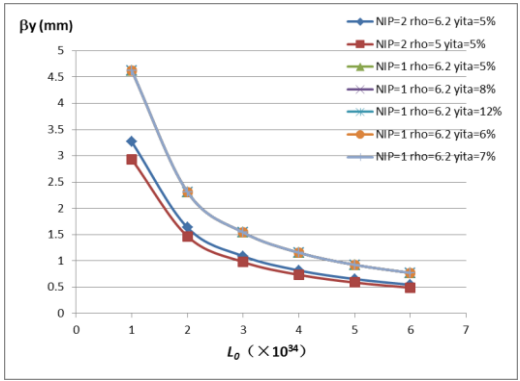

FIG. 1. Vertical beta at IP as the function of peak luminosity

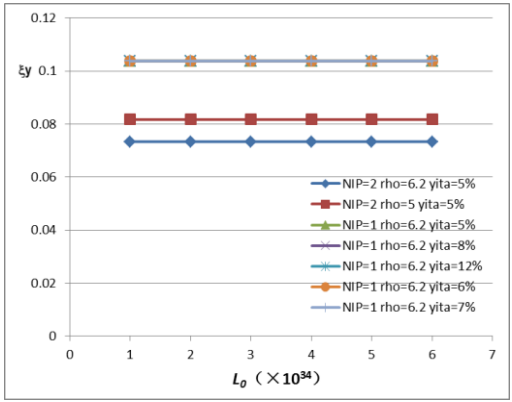

FIG. 2. Vertical beam-beam tune shift as the function of peak luminosity

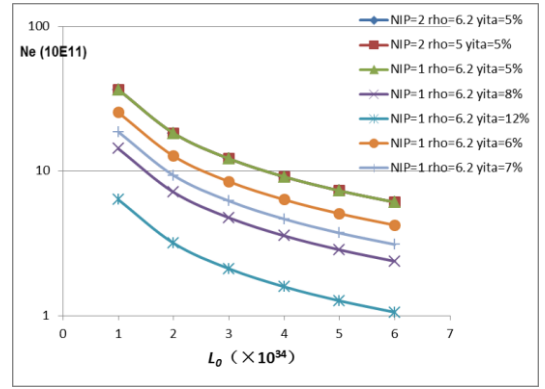

FIG. 3. Bunch population as the function of peak luminosity

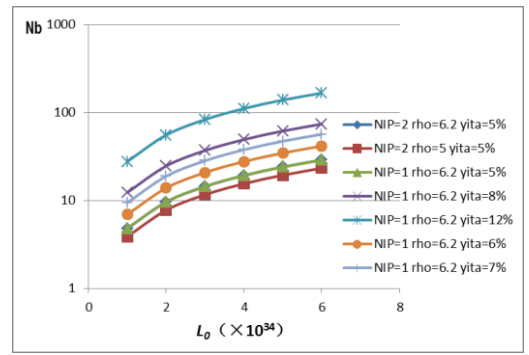

FIG. 4. Bunch number as the function of peak luminosity

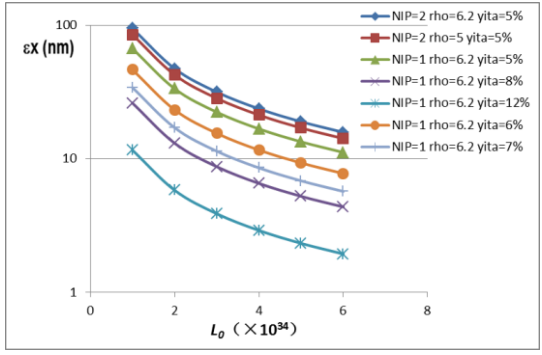

FIG. 5. Horizontal emittance as the function of peak luminosity

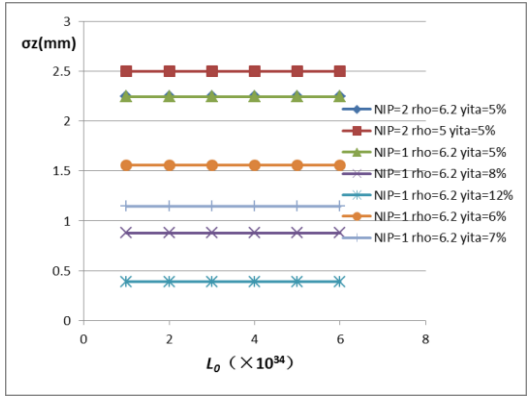

FIG. 6. Bunch length as the function of peak luminosity

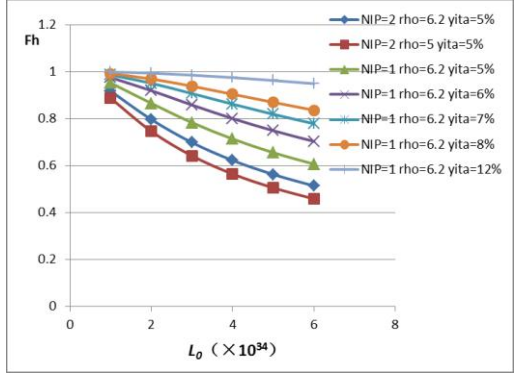

FIG. 7. Hour glass factor as the function of peak luminosity

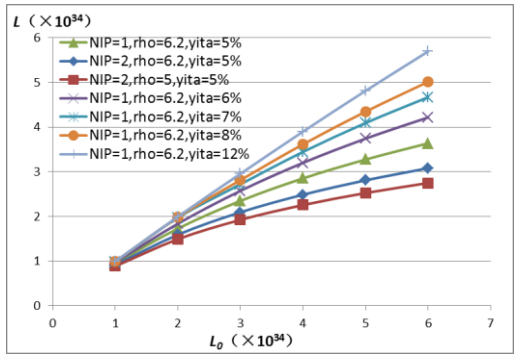

FIG. 8. Real luminosity as the function of peak luminosity

Overall speaking, we should decrease IP number, increase bending radius and energy acceptance in order to achieve higher luminosity. Obviously $N_{IP}=1$ is the minimum value for IP number. Assuming the maximum fill factor of the dipoles is 80%, 6.2 km bending radius will be a limit for the 50 km ring. Then what we need to consider about is how large the energy acceptance can reach and which parameter constraints the enlargement of energy acceptance.

### 5.2. Constraints from RF system

As long as a set of beam parameters is determined, we need to check the RF system to see if the bunch length and energy acceptance can be achieved.

Firstly, considering the synchrotron radiation energy loss have to be compensated by the RF cavities, one finds

$$U_0 = eV_{rf} \sin\phi_s \qquad (41)$$

where $V_{rf}$ is the total voltage for the RF cavities and $\phi_s$ is the synchrotron phase. According to eq. (41), one gets

$$\phi_s = \pi - \arcsin\left(\frac{U_0}{eV_{rf}}\right) \qquad (42)$$

The nature bunch length is expressed by

$$\sigma_z = \sqrt{-\frac{2\pi E_0 \alpha_p}{f_{rf} T_0 eV_{rf} \cos\phi_s}} \bar{R}\delta_0 \qquad (43)$$

where $\alpha_p$ is the momentum compaction factor, $f_{rf}$ is the RF frequency and $\bar{R}$ is the average radius of the ring. Then, the expression for the energy acceptance is

$$\eta = \sqrt{\frac{2U_0}{\pi\alpha_p f_{rf} T_0 E_0}\left(\sqrt{q^2-1} - \arccos(\frac{1}{q})\right)} \qquad (44)$$

where $q = \frac{eV_{rf}}{U_0}$. Combining the eqs. (43) and (44), we can get the RF frequency $f_{rf}$ and the momentum compaction $\alpha_p$ for given RF voltage $V_{rf}$ and energy acceptance $\eta$.

In order to see how large the energy acceptance we can get, we make a scan of energy acceptance with different RF voltage (The bending radius is fixed to be $\rho$=6.2 km). The results are shown in Fig. 9 and Fig. 10.

From Fig. 9, one finds that larger energy acceptance need higher RF frequency and lower RF voltage indicates lower RF frequency for the fixed energy acceptance. Fig. 9 shows a linear dependence of the RF frequency to the energy acceptance. If one wants to choose 350 MHz RF frequency like LEP2 the corresponding energy acceptance is about 3%, and if one prefers 1.3GHz RF technology the energy acceptance will be about 8%. In other words, the maximum luminosity which we can obtain is closely related with the RF technology (frequency). From the beam dynamics point of view, lower RF frequency is a better choice because the cavities with lower frequency have larger aperture and hence lower impendence which is a favor for the collective instabilities. Also considering there are still technical difficulties to directly use ILC 1.3 GHz SC technology on storage rings [12], it's better to choose the frequency lower than 1GHz (700 MHz for example).

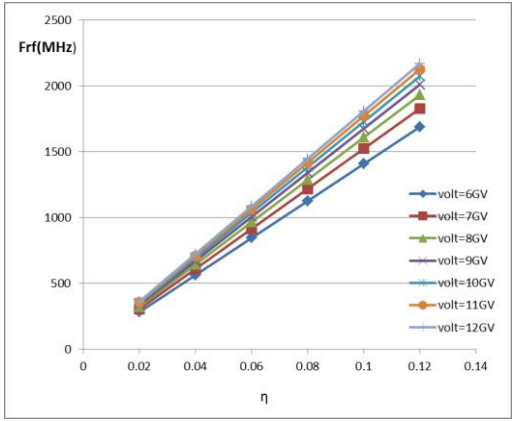

FIG. 9. RF frequency as the function of energy acceptance

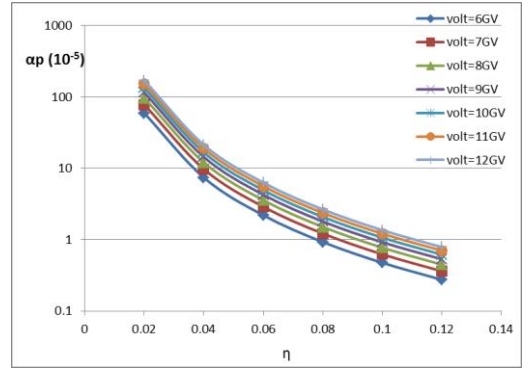

FIG. 10. Momentum compaction factor as the function of energy acceptance

From Fig. 10, we can see that the requirement of enlarging energy acceptance is translated to design a low momentum compaction lattice and also larger RF voltage will relax this tolerance and lose the difficulties of lattice design. So we need to make a reasonable choice for the total RF voltage while balancing the constraints from the RF frequency and momentum compaction.

### 5.3. Optimized machine parameters

Combining the discussions in 5.1 and 5.2, we get a set of new designs for the 50 km Circular Higgs Factory with three typical RF frequencies corresponding to different RF technology (Table 2). For these designs, we choose $\rho$=6.2 km to get the maximum luminosity and each time the peak luminosity $L_0$ is raised to a highest value until the minimum $\beta_y$ (confine $\beta_y$ at IP will not smaller than 1 mm) is reached.

TABLE II. Optimized parameters of Circular Higgs Factory (CHF) with different RF technology

|  | *350 MHz (LEP2-like) technology* | *700 MHz technology* | | *1.3 GHz (LEP3-like) technology* |
|---|---|---|---|---|
| Number of IPs | 1 | 1 | 2 | 1 |
| Energy (GeV) | 120 | 120 | 120 | 120 |
| Circumference (km) | 50 | 50 | 50 | 50 |
| SR loss/turn (GeV) | 2.96 | 2.96 | 2.96 | 2.96 |
| $N_e$/bunch ($10^{12}$) | 1.61 | 0.79 | 1.12 | 0.33 |
| Bunch number | 11 | 22 | 16 | 53 |
| Beam current (mA) | 16.9 | 16.9 | 16.9 | 16.9 |
| SR power /beam (MW) | 50 | 50 | 50 | 50 |
| $B_0$ (T) | 0.065 | 0.065 | 0.065 | 0.065 |
| Bending radius (km) | 6.2 | 6.2 | 6.2 | 6.2 |
| Momentum compaction | 0.43 | 0.38 | 0.38 | 0.21 |
| (10$^{-4}$) | | | | |
| $\beta_{IP}$ x/y (m) | 0.2/0.001 | 0.2/0.001 | 0.2/0.001 | 0.2/0.001 |
| Emittance x/y (nm) | 29.7/0.15 | 14.6/0.073 | 29.1/0.15 | 6.1/0.03 |
| Transverse $\sigma_{IP}$ (um) | 77/0.38 | 54/0.27 | 76/0.38 | 35/0.17 |
| $\xi_x$/IP | 0.103 | 0.103 | 0.073 | 0.103 |
| $\xi_y$/IP | 0.103 | 0.103 | 0.073 | 0.103 |
| $V_{RF}$ (GV) | 4.1 | 6 | 6 | 9.3 |
| $f_{RF}$ (MHz) | 350 | 704 | 704 | 1304 |
| $\sigma_z$ (mm) | 4.6 | 2.2 | 2.2 | 0.95 |
| Energy spread (%) | 0.13 | 0.13 | 0.13 | 0.13 |
| Energy acceptance (%) | 3.5 | 5 | 5 | 7.7 |
| $\gamma_{BS}$ (10$^{-4}$) | 9.7 | 13.8 | 13.8 | 21.3 |
| $n_\gamma$ | 0.86 | 0.6 | 0.6 | 0.39 |
| $\delta_{BS}$ (10$^{-4}$) | 4.3 | 4.3 | 4.3 | 4.3 |
| Life time due to beamstrahlung (minute) | 30 | 30 | 30 | 30 |
| F (hour glass) | 0.49 | 0.68 | 0.68 | 0.87 |
| $L_{max}$/IP (10$^{34}$cm$^{-2}$s$^{-1}$) | 2.2 | 3.1 | 2.2 | 4.0 |

## 6. Conclusion

In this paper, a general method of how to make an optimized machine parameter design of a circular e+e- Higgs Factory by using analytical expression of maximum beam-beam tune shift and beamstrahlung beam lifetime started from given luminosity goal, beam energy and technical limitations was developed. By using this method, one reveals the relations of machine parameters with goal luminosity clearly and hence give an optimized design in an efficient way. Also, we point out that the highest luminosity which we can get is closely related with the RF technology (frequency) and higher luminosity favors higher RF frequency. So the maximum luminosity that is realizable is subject to the detail RF technology. Finally a series of optimized designs with different RF frequency for 50 km Circular Higgs Factory was proposed based on beam parameters scan and RF parameters scan. Up to now, the luminosity we got is the highest one among the exist designs.

**Acknowledgments**

The authors would like to thank the support and suggestions from Professor Yifang Wang. This work was supported by the National Foundation of Natural Sciences Contract 11175192.